\documentclass[11pt,twoside]{article}

\usepackage{asp2004}
\usepackage{epsf}
\usepackage{psfig}
\usepackage{lscape}

\begin{document}
\title{Formation and Evolution of Planetary Systems:  Placing Our Solar System in Context with Spitzer}   
\author{Michael R. Meyer}   
\affil{Steward Observatory, The University of Arizona} 
\author{and the FEPS Legacy Science Team} 
\affil{http://feps.as.arizona.edu}    

\begin{abstract} 
We summarize the progress to date of
our Legacy Science 
Program entitled {\it The Formation and Evolution of Planetary
Systems} (FEPS) based on observations obtained with the Spitzer Space 
Telescope during its first year of operation.  In addition to 
results obtained from our ground--based preparatory program 
and our early validation program, we describe new results from a survey for
near--infrared excess emission from the youngest stars in our
sample as well as a search for cold debris disks around sun--like
stars.  We discuss the implications of our findings with respect
to current understanding of the formation and evolution of
our own solar system.
\end{abstract}


\section{Introduction}   

Is our solar system, and the potential for life to arise that it 
represents, common or rare among sun--like stars in the disk of 
the Milky Way galaxy?  In order to help answer this question, 
planetary scientists use the properties observed today
in our solar system to infer its evolutionary history and 
initial conditions of formation.  Astronomers have a different
approach, observing sun--like stars at a
variety of ages in an attempt to constrain the 
range of paths taken by evolving planetary systems that might
surround these stars.  Only by synthesizing results from 
studying our solar system and observations of disk evolution
around ensembles of sun--like stars can we address the question posed. 

Few researchers dispute that most stars are surrounded by 
circumstellar accretion disks at birth.  Recent near--infrared (1--4 $\mu$m) 
photometric studies have 
indicated that this accretion phase lasts between 1--10 Myr
with 50 \% of sun--like stars losing their inner accretion
disks (R$_{disk} <$ 0.1 AU) within 3 Myr (Haisch et al. 2001). 
Yet the gas and dust content of 
remnant disks at larger radii is unconstrained by these 
observations.  Cooler dust at larger radii will emit at 
longer wavelengths characteristic of the appropriate
blackbody temperature (Beckwith, 1999).  As a result, 
we can consider  that 
different wavelengths trace different radii in the disk 
It is extremely important to conduct photometric studies 
from 1--1000 $\mu$m as a function of age 
in order to constrain the geometric 
distribution of dust from 0.1--100 AU and how it may
evolve over time.   

Observations of accretion disks
surrounding T Tauri stars have established that they 
are optically--thick from a few to $>$ 30 $\mu$m (Beckwith et al. 1990; 
see also D'Alessio et al. 2001). 
It is from these disks that we think planets form. 
As they do, material in the disk is either incorporated
into larger objects or dispersed leading to the optically--thin 
disks observed around older main sequence stars. 
How much material is required for optical--depth 
unity perpendicular to the disk?  If the disk interior 
to 0.1 AU were optically--thick and the opacity were
dominated by gas, that would imply a mass accretion
rate $>$ 10$^{-7}$ M$_{\odot}$/yr (Meyer et al. 1997; 
see also Muzerolle et al. 2003).  If the opacity is
dominated by dust, it would take roughly a few times 
the mass of the asteroid Ceres distributed interior to 
0.1 AU in micron--sized particles to reach optical
depth unity.  In the mid--infrared, it would take approximately
one Earth mass of micron--sized grains between 0.1-1 AU
for the dust to be optically--thick.  And in the far--infrared,
approximately one Jupiter's worth of small grains distributed
between 1--10 AU would correspond to $\tau=1.0$. 

It is often assumed that optically thin disks correspond
to debris disks rather than primordial disks, 
though this need not be the case.  What do we mean by these
distinctions and how can we discern them?  
We will refer to a disk as {\it primordial} if the opacity
in the disk is dominated by interstellar dust grains incorporated
into the disk as a by--product of the star formation process.
We will refer to a disk as {\it debris} if the opacity is
dominated by grains generated through collisions of larger
parent bodies.  While these definitions are helpful in clarifying
the terminology, they are not very practical.   We might hope
to tell the difference observationally by constraining 
the amount of remnant gas in a disk, thereby invoking timescale
arguements (P--R drag and radiation pressure blowout) to 
infer that the dust we see must be continuously replenished
(Backman and Paresce, 1993) or detecting
solid--state spectral features that suggest dust generated
from a differentiated parent body. 

In what follows, we review some recent results concerning
the evolution of circumstellar disks surrounding sun--like stars
in the pre--Spitzer era, describe briefly our Legacy Science Program, 
provide an overview of our initial results, and discuss their 
implications.  Our main goals in carrying out this program are
to constrain theories for the formation and evolution of 
planetary systems, and to provide insight into 
whether solar systems like our own are common or rare surrounding
sun--like stars in the disk of the Milky Way. 

\section{The Pre--Spitzer Era}   

While the IRAS satellite provided powerful evidence for circumstellar
accretion disks surrounding newly formed stars (Rucinski et al. 1985) as well
as debris disks surrounding nearby luminous main sequence stars (Aumann, 
1985), many gaps remain in our understanding of how to connect these
two phenomena.  Several studies undertaken with the Infrared Space
Observatory (ISO), have helped make these connections 
(see review by Meyer and Beckwith, 2000; 
van Dishoeck, 2003) and focus attention on 
unresolved questions:  1)  do debris disks diminish with the mass 
in small grains as $t^{-2}$ expected if the dominant removal 
mechanism for grains is P--R drag (e.g. Spangler et al., 2001;
see however Decin and Dominik 2003); 2) how long do gas--rich disks persist
in circumstellar disks (e.g. Thi et al. 2001); 3) can we infer
the presence of massive planets in disks based on spectral energy
distributions (SEDs) (e.g. Bouwman et al. 2003); and 
4) can we trace the evolution of grain size and dust composition
through observations of solid state features in the remnant dust
(e.g. Meeus et al. 2001).  

Recent ground--based studies undertaken
in preparation for the Spitzer Space Telescope have also furthered
our understanding.  Mamajek et al. (2004; see also Weinberger et al., 
2003 and Metchev et al., 2004) 
conducted sensitive mid--infrared photometric studies 
in order to constrain the evolution of dust in the terrestrial 
planet zone.  These preliminary results suggest
that optically--thick disks between 0.1--1.0 AU evolve on timescales 
comparable to the cessation of accretion.  Further, the limits placed on 
the amount of optically--thin debris disks appear inconsistent
with simple extrapolations of our own solar system zodiacal dust
disk to ages of 10--30 Myr (Mamajek et al. 2004).  Sub--millimeter
surveys for cool dust surrounding sun--like stars (e.g. Carpenter
et al. 2005) provide quantitative constraints on the evolution
of dust mass from 3--30 Myr.  Following up a rare dust detection
towards the 100--300 Myr old G star HD 107146, Williams et al. 
(2004) tentatively resolved the 450 $\mu$m emission and modelled
the disk with a large inner hole.  Ardila et al. (2005) have 
confirmed aspects of this model with images of the disk in 
scattered light with the ACS on HST.  Finally, Kessler et al. 
(2005) describe ground--based efforts to extend work done on 
evolution of the silicate emission features observed 
toward Herbig Ae/Be stars with ISO (Meeus et al. 2001) 
to sun--like stars from the protostellar phase through to the 
debris disk phase. 

\section{Overview of the FEPS Program}   

In coordination with guaranteed time programs, general observer programs, 
and other Legacy Science Programs, the FEPS project
attempts to address: 1) how optically--thick accretion disks
transition to become debris disk systems; 2) when the molecular
gas from which giant planets form is removed from circumstellar
disks around sun--like stars; and 3) the diversity of planetary
architectures inferred from infrared observations of main sequence stars. 
Our sample consists of approximately 300 stars with masses between 
0.7--1.5 M$_{\odot}$ spanning a range of ages from 3 Myr to 3 Gyr. 
Roughly 50 stars were selected in six equally spaced bins of log--age.
Care was taken to choose the closest targets available in the 
lowest IR background regions from the parent samples available
within the constraints of our program.  Each of our target
stars is observed with every Spitzer instrument: IRAC 
photometry in the near--infrared, IRS low resolution spectroscopy 
in the mid--infrared, and MIPS photometry in the far--infrared. 
A sub--set of our sample is observed with the IRS at high 
spectral resolution in order to discern the amount of remnant
gas in disks through a variety of atomic and molecular species
(see Hollenbach et al., this volume).   Additional details concerning the 
FEPS program can be found in Meyer et al. (2005). 

\section{New Results}   

We received our first Spitzer data in December, 2003, as part of
the Legacy Validation Program in order to insure that our observing
proceedures were suitable for our program given the measured 
in--flight performance and modified operation protocols.  Although 
minor modifications were made to the overall program based on these
data, we succeeded in: 1) confirming the presence of a debris disk 
surrounding the 30 Myr old sun--like star HD 105, 
2) discovering a new debris disk surrounding HD 150706, a G star
with age 700 $\pm$ 300 Myr, and 
3) placing stringent upper limits on dust mass in three older systems. 
For the two debris disks, 
inner holes ranging from 20--45 AU were inferred based on the
observations.  The limits on the amount of warm debris 
suggest that the dust surface density drops by x100 inside
the inner hole radii.  While these inner holes could be maintained
collisionally, depending on the assumptions made concerning the 
dust, they are also consistent with a model requiring the presence 
of gas giant planets (Moro--Martin et al., 2005). 

Several other studies are underway based on data collected during the 
first year of Spitzer operations, some of which was delivered to the 
Spitzer Legacy Science Archive in October, 2004.  In a follow--up 
study concerning the lifetime of inner dust disks surrounding young
stars, Silverstone et al. (2005) confirm and extend the results 
from Mamajek et al. (2004).  In a sample of 74 stars ranging in 
age from 3--30 Myr, only five stars were found to exhibit emission
in excess of that expected from the stellar photosphere at wavelengths
of 3.6, 4.5, and 8.0 $\mu$m.  While four of the excess stars have ages
from 3--10 Myr, only one star with excess was found with an age from 
10--30 Myr.  In all cases the excesses arise from optically--thick 
disks with $4/5$ stars exhibiting spectroscopic evidence for 
active disk accretion.
No stars in the sample were found to have optically--thin excess 
emission with constraints of $<$ 10$^{-5}$ Earth masses of dust in 
micron--sized grains inferred for non--detections.  This suggests that:
1) dust from 0.1--1.0 becomes extremely optically--thin on timescales of 
10 Myr; and 2) the timescale to transition between optically--thick and
thin is $<$ a million years.  

\begin{figure}[!ht]
\plotone{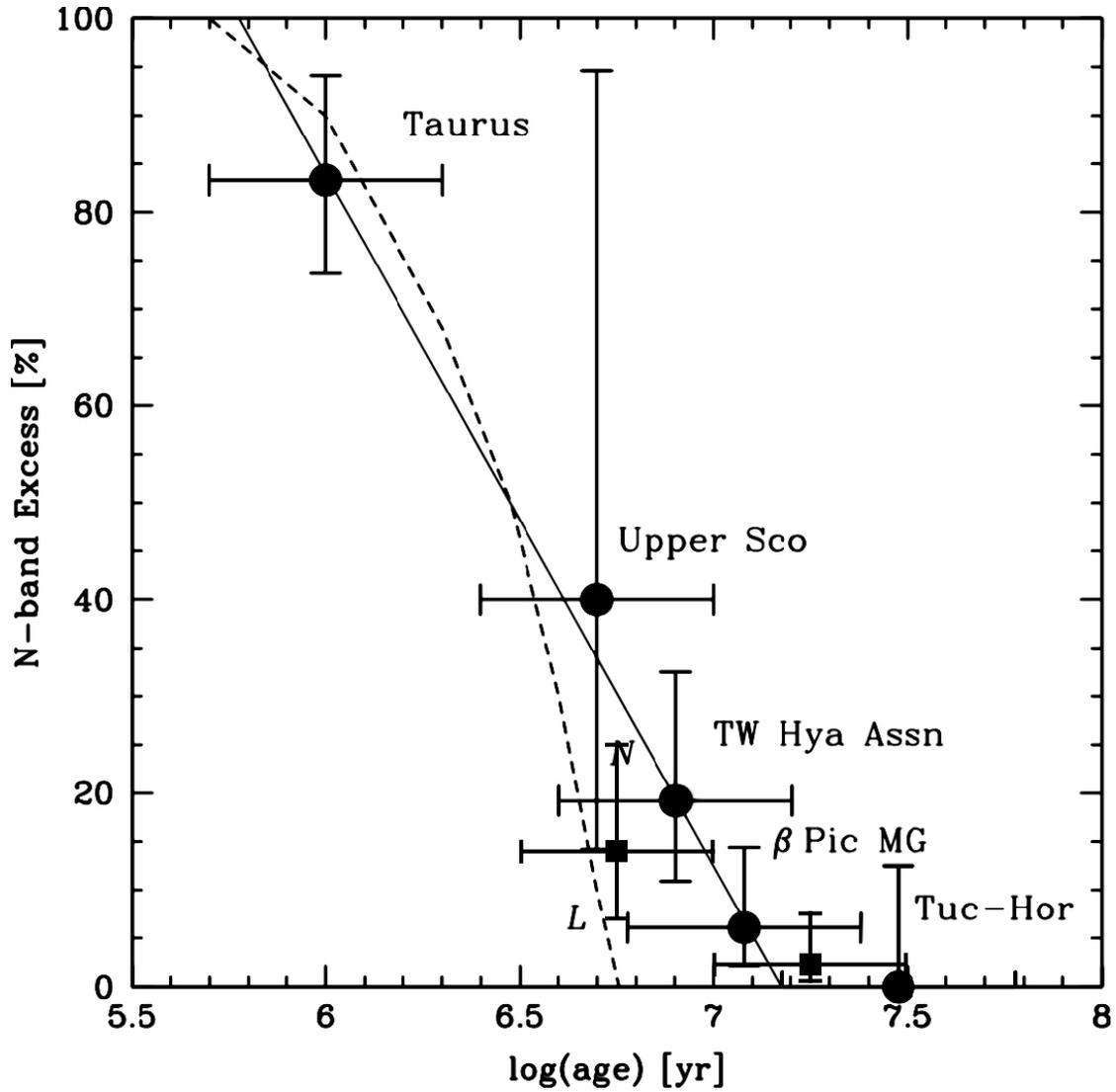}
\caption{
Frequency of stars in clusters or associations
with detectable mid--infrared excess versus cluster age.
Filled circles are data taken from the literature while the filled
squares are new results from Silverstone et al. (2005).}
\end{figure}

Hollenbach et al. (this volume; 2005) describe the search for remnant
gas using the IRS high resolution mode to search for emission
lines of atomic and molecular species expected to dominate the
cooling of gas disks.  Non--detections towards the 30 Myr debris
disk candidate HD 105 suggest upper limits of $<$ 0.1 M$_{JUPITER}$
in remnant gas, ruling out the future formation of gas giant planets
in that system.  Hines et al. (this volume; 2005) describes
the search for warm debris disks detected with the IRS including
the unusual ring of material surrounding HD 12039 emitting at 110 K.  
Assuming the emission comes from large blackbody grains, models suggest
that the material is confined to a narrow ring from 4--6 AU
not unlike the asteroid belt in our own solar system. 

Kim et al. (2005) present new data for five cold debris disks 
surrounding sun--like stars analogous to the dust generated in our solar 
system by collisions of Kuiper Belt objects.  In all cases, limits
are placed on the amount of warm inner debris and estimates are made
for the sizes of the inner disk holes inferred as a function of assumptions
made concerning the grain properties.  There is a 
degeneracy in the models such that the SEDs
observed can be explained either through 
large (blackbody) dust grains at radii from 20--50 AU, or 
small grains near the blowout size at radii $>$ 100 AU.  
Minimum dust masses inferred are consistent (within a factor of x3) 
with extrapolations based on a toy model for the evolution
of the Kuiper Belt (Backman et al. 2005).
While models where the inner holes are maintained by collisional 
evolution and radiation pressure blowout cannot be ruled out, 
the data are also consistent with inner holes 
maintained by the presence of gas giant planets.
We plan to distinguish between these models by: 1) attempting to 
directly detect gas giant planets surrounding the youngest and
nearest cold debris disks in our sample (e.g. Masciadri et al., 2005), 
2) coronagraphic imaging of the outer dust disks in scattered light 
(e.g. Kalas et al. 2005), and 3) resolved
observations of the thermal emission using nulling interferometry
on large ground--based telescopes equipped with adaptive optics
(e.g. Liu et al. 2004). 

\begin{figure}[!ht] 
\plotone{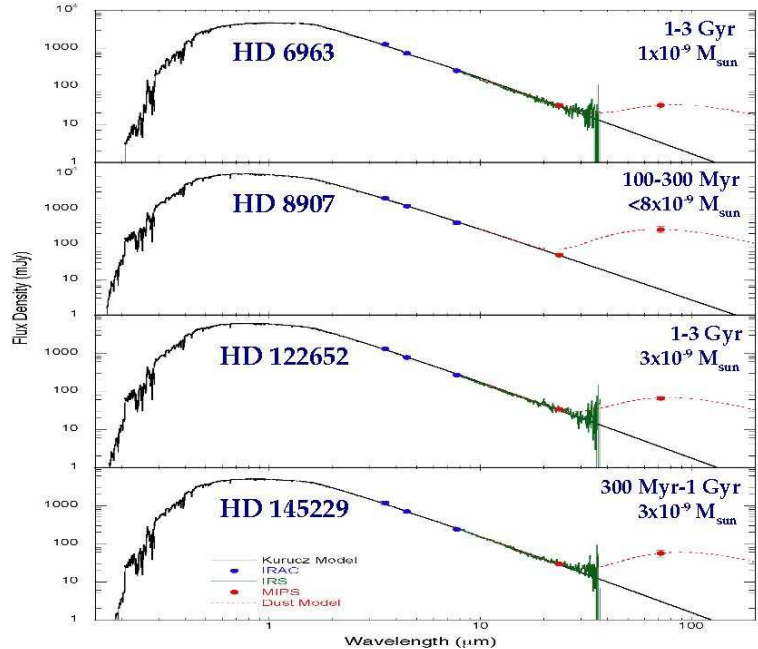}
\caption{Spitzer SEDs for 
debris disks surrounding sun--like stars from Kim et al. 
(2005).  Expected
photospheric emission based on our model fits are shown
as solid lines, with best fit models for the excess emission 
indicated with a dashed line.  The ages and estimated dust masses 
for each system are also given.} 
\end{figure}

\section{Implications}   

What sort of picture is emerging concerning the formation
and evolution of planetary systems around sun--like stars
based on the early results from FEPS?  Our results 
suggest that not only are inner disks surrounding
post--accretion T Tauri stars devoid of detectable disks out
to a few AU, but also that the dust production rates in any
remnant planetesimal belts are lower than predicted by simple
models for the early evolution of our solar system. 
Whether these
systems lack planetesimal belts all together, or have yet to be
dynamically stirred by the presence of larger bodies remains to 
be seen.  We cannot yet distinguish whether outer debris disks
are best fit with a diminution of dust mass as t$^{-2}$ or t$^{-1}$
characterisic of P--R drag and collisional blowout dust removal 
mechanisms, respectively.  What is clear is that there is a large 
dispersion in dust masses at all ages.  Rieke et al. (2005) present
evidence that the typical disk evolution around intermediate--mass 
(A--type) stars could be
punctuated by rare collisions that temporarily raise observed dust 
signatures by orders of magnitude. 

\begin{figure}[!ht] 
\plotone{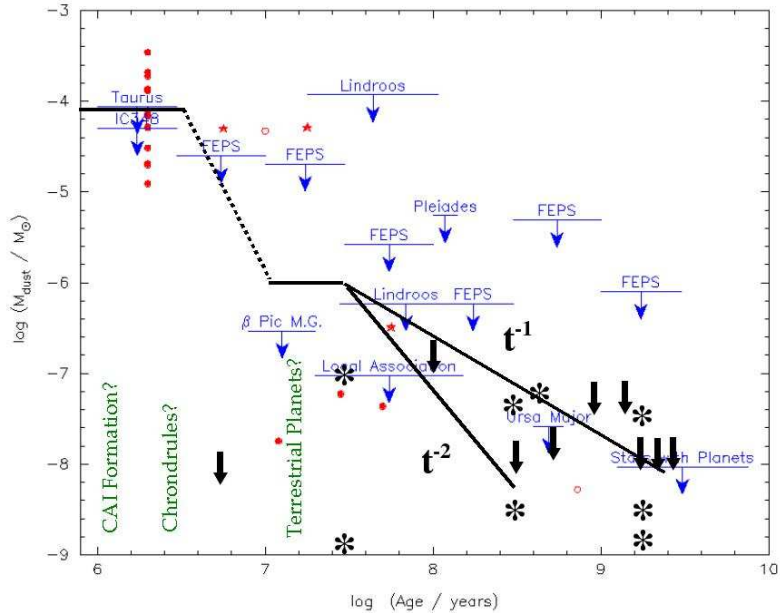}
\caption{Dust mass versus age with sub--mm 
observations of Carpenter et al. (2005) compared with 
70 $\mu$m detections (and upper limits) from Meyer et al. (2004)
and Kim et al. (2005).  Also shown are two 
models for dust diminution with time.  Evolution with $t^{-2}$ 
represents the P--R drag model while $t^{-1}$ indicates collisional 
blowout removal.} 
\end{figure}

Future work includes: 1) a study of dust excess surrounding G stars
in the 100 Myr old Pleiades open cluster (Stauffer et al.), 2) the
properties of warm debris disks in our FEPS sample (Bouwman et al.), 
3) the spectral energy distributions and solid state spectra 
of long--lived accretion disks uncovered in our sample (Bouwman et al.), 
and 4) upper limits to the lifetime of gas--rich disks based on our
survey of 10 stars with the IRS in high resolution mode (Pascucci et al.). 
Of paramount importance will be development of a model for the dust
production rates in our own solar system as a function of time so 
that we can ascertain whether our solar system is consistent with 
the observed dispersion in evolutionary properties for disks surrounding
sun--like stars. 
In this way, we hope to address one piece of the puzzle in 
understanding whether solar systems like our own are common
or rare in the Milky Way. 

\acknowledgements 

We thank the organizers of this conference, especially Lee Armus, 
 for the opportunity to present our work.  We are gratefully for
the help and advice from our colleagues at the Spitzer Science 
Center and the IRAC, IRS, and MIPS instrument teams.  We gratefully
acknowledge support from NASA grants 1224768, 1224634, and 1224566 
administered through JPL. 


\end{document}